\documentclass[11pt]{article}
\usepackage{graphicx}

\newcommand\pubnumber{TTP17-058}
\newcommand\pubdate{\today}

\def\napoli{Institute for Theoretical Particle Physics\\
Karlsurhe Institute of Technology, Karlsruhe, Germany}

\def\Title#1{\begin{center} {\Large #1 } \end{center}}
\def\Author#1{\begin{center}{ \sc #1} \end{center}}
\def\Address#1{\begin{center}{ \it #1} \end{center}}

\newcommand\pubblock{\rightline{\begin{tabular}{l} \pubnumber\\
         \pubdate  \end{tabular}}}
\newenvironment{Abstract}{\begin{quotation}  }{\end{quotation}}
\newenvironment{Presented}{\begin{quotation} \begin{center} 
             PRESENTED AT\end{center}\bigskip 
      \begin{center}\begin{large}}{\end{large}\end{center} \end{quotation}}
\def\Acknowledgements{\bigskip  \bigskip \begin{center} \begin{large}
             \bf ACKNOWLEDGEMENTS \end{large}\end{center}}
\def\beq{\begin{equation}}
\def\eeq#1{\label{#1}\end{equation}}
\def\eeqn{\end{equation}}

\def\beqa{\begin{eqnarray}}
\def\eeqa#1{\label{#1}\end{eqnarray}}
\def\eeqan{\end{eqnarray}}

\let\bar=\overbar

\def\Dslash{\not{\hbox{\kern-4pt $D$}}}
\def\dslash{\not{\hbox{\kern-2pt $\del$}}}

\def\msb{{\bar{\ssstyle M \kern -1pt S}}}

\begin{document}
\begin{titlepage}
\pubblock

\vfill
\Title{TOP~2017: Theoretical Summary}
\vfill
\Author{Kirill Melnikov}
\Address{\napoli}
\vfill
\begin{Abstract}
I summarize the theory talks presented at the TOP~2017 conference.
\end{Abstract}
\vfill
\begin{Presented}
10$^{th}$ International Workshop on Top Quark Physics TOP~2017 \\
Braga, Portugal,  September 17--22, 2017
\end{Presented}
\vfill
\end{titlepage}
\def\thefootnote{\fnsymbol{footnote}}
\setcounter{footnote}{0}

\section{Introduction}

In contrast to almost all other participants of  the TOP~2017 conference, 
I do not actively work on the 
top quark physics these days.  Indeed, my last paper 
on top quark physics was published more than three years ago. Given 
the task of summarizing the theory talks at the TOP~2017 conference, 
there is both 
bad and good   in this  situation: bad, 
because my understanding and knowledge of  certain things  can  be 
incomplete or  outdated; good, because it becomes easier to notice   
and appreciate new developments that occurred  during elapsed time 
and to reflect  on the changes from a broader perspective.  

In this respect, the week of TOP 2017 in 
Portugal offered many interesting observations  that 
collectively point towards 
truly exciting  progress  
that is happening    in the top quark world. On the experimental 
side, it is the appearance of very large data samples that, potentially, 
offer many  hidden gems 
if one knows where to look and which questions to ask. 
On the theory 
side, it is an emergence of high-precision predictions for 
many different  processes 
with top quarks  and a growing appreciation that many 
fundamental things can be  learned by utilizing them in the right way. 

These changes are driven by  a lively community of physicists 
who have  common goals, talk 
the same language and appreciate each others contributions to the exploration 
of the top quark physics. 
The close cooperation of theory and experiment 
leads to many  interesting results, beyond 
traditional exclusion limits,  that are based 
on a careful theoretical  interpretation of precise measurements. 
These types of interactions seem to be the 
hallmark of the top quark physics but  
they can also  be   viewed as a model for how LHC 
physics will be done  in the years to come. If TOP~2017 is  
any indication,  high energy physics future is assured  since 
there will be non-trivial problems to solve, 
interesting discussions to have  and ambitious scientific goals to achieve, 
 even if direct production of new particles at the LHC will remain out of reach. 

The  vitality of  the top quark physics  could be felt especially 
strongly at TOP~2017 since it offered a stark 
contrast to a somewhat gloomy perspective on the future 
of high energy physics driven by the fact that no BSM physics 
has been  discovered so far at the LHC.  The question that looms 
large over the horizon  is how to move forward from here.  
One of the  possible answers -- if not {\it the only} possible answer 
-- is that we  have to start serious discussions   about credible  and verifiable 
ways to   find  subtle effects and small deviations from Standard Model 
predictions at hadron 
colliders or, said differently,  about how to do 
precision physics at the LHC in a convincing way.  
From this perspective, a particular relevance 
of the LHC top quark physics program is that it has never been  entirely 
focused on searches,   
since it {\it anyhow}  had a new heavy particle  to study,  
and studying this  particle was, in fact, what was being done.  
As the result,  when  searches turned  null 
results, the ``precision mentality'' was 
already in place in the top quark community 
and it was ready to take the lead.

Precision physics goes after  phenomena that are not easy to see 
in hadron collisions. They  may arise 
in many ways as  their originators can be too heavy to be seen or they 
can blend into a background.  No matter how subtle, one hopes to find a way 
to understand  what they are.  The success of 
this endeavor requires several  things: i) superb experimentation;
ii) good understanding of which  questions to 
ask and a good eye for places where new things can hide and 
iii) an 
ability to describe hadron collisions from first principles with maximal 
attainable (and still sensible) precision.   The cross-talk between 
experts in different theory areas and experts in experiment is crucial; 
it is precisely  this cross talk that will allow us to move forward towards the common 
goal -- finding physics beyond the Standard Model at the LHC either directly 
or through precision measurements.  It was 
inspiring to see how efficiently 
this cross talk works in the top quark community. In fact, it appears that 
the ability to connect very different aspects of high-energy physics 
is driven by the very nature of the top quark, a particle that 
has something for everyone. Indeed, 
i)  it is unusually heavy and interacts  strongly with the 
Higgs boson; so strongly, in fact, that it can destabilize the 
electroweak vacuum \cite{craig,salvio}; 
ii) it is part of a flavor puzzle but the role it plays is  not apparent
\cite{craig,panico, takeuchi};
iii) it may be expected to talk directly to the Dark Side, but we do not 
know how and if at all
\cite{craig,kilic}; 
iv) it has a capacity to annoy those of us who do not care about the top 
by directly interfering with searches for other, ``more interesting'' things
at the LHC; 
v) it is the only ``free'' colored  particle that we can observe and whose 
properties we, therefore, can describe in great detail from first principles. 
In short, top quark is an interesting object to study and it  can teach us a lot. 
This was the leitmotif of the many theory talks at the TOP~2017 conference that I attempt 
to summarize in the remainder of this contribution. 
% In the remainder 
%of this contribution, I summarize  the theory talks at the 
%TOP~2017 conference.

%%%%%%%%%%%%%%%%%%%%%%%%%%%%%%%%%%%%%%%%%%%%%%%%%%%%%%%%%%%%%%%%%%%%%%%%%
%%
%%   use this format to include an .eps figure into your paper
%%
\begin{figure}[t]
\centering
\includegraphics[trim = 12 300 15 300, clip,  width=.7\linewidth]{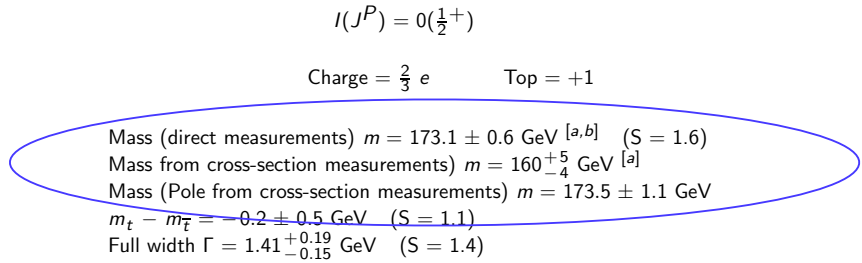}
\caption{Top quark quantum numbers and  three (!) masses from Review of Particle Physics.}
\label{fig:fig1}
\end{figure}
%%%%%%%%%%%%%%%%%%%%%%%%%%%%%%%%%%%%%%%%%%%%%%%%%%%%%%%%%%%%%%%%%%%%%%%%%%%

\section{The top quark mass}
How many parameters  identify the top quark and define 
its physics? This is an interesting question to discuss~\cite{craig}
but, if in doubt,  we can always  consult the  
most learned resource  that we have in our  field  
-- the Review of Particle Physics. 
The result is shown in Fig.~\ref{fig:fig1} 
which  reminds  us about the  electric 
charge of the  top quark, its weak isospin and  ``topness'', 
gives   the  value of 
the top  quark width and, amazingly  
enough, lists {\it three} different top quark masses.
Three masses  is a puzzling feature 
as it appears  that   the top quark 
has three different {\it types} of masses  
that go under the following 
names: 1) mass  from direct 
measurement, 2) mass from cross section measurements and 
3) pole mass from cross sections measurements.  
Clearly, for those who still do not believe that the top quark is special, 
this is a very strong argument   that in certain  aspects 
this particle is truly unique!

Of course, ``three top quark masses'' is an indication of a problem that 
we run into  when trying to understand what this important quantity means. 
For the sake of example, it is often said that 
the value of the top quark mass has important implications 
for the stability of electroweak vacuum. 
It is stated that the electroweak vacuum becomes 
stable if the {\it pole mass} of the top 
quark $m_t$ is smaller  than  $171.18~{\rm GeV}$. Since 
the numerical value for $m_t$ reported by 
ATLAS and CMS is $172.4(5)~{\rm GeV}$, it appears 
to be precise enough to argue that Universe is not stable and that its
lifetime is  $10^{139^{+105}_{-51}}$ years~\cite{craig}. 
However, the problem 
with these discussions  is that in both of these statements the mass parameter 
whose numerical value is quoted is not specified!  This omission 
could be excused if not for the fact that the most natural definition 
of a mass parameter 
of a heavy particle -- the pole mass -- is ambiguous for a heavy 
quark by an amount proportional to 
 $\Lambda_{\rm QCD} \sim {\cal O}(150-200)~{\rm MeV}$.
Moreover, although the experimental number is very precise, it is often 
questioned if it corresponds to the pole mass or to some other mass 
parameter such as e.g.  the ``Monte Carlo''  mass. 

Both of these issues  were discussed at the conference by 
P.~Nason~\cite{nason}.  He argued that the theoretical 
uncertainty in the pole  mass of the top quark related to 
the asymptotic nature of perturbative series  (infra-red 
renormalon) 
is  ${\cal O}(100-200)~{\rm MeV}$~\cite{nason} and, therefore, is  much 
smaller than the current experimental uncertainty on the top quark mass. 
Given that the error on the measurement is ${\cal O}(500)~{\rm MeV}$ 
and that it is unlikely to decrease significantly, the irreducible 
theory  uncertainty in the value of the top quark pole 
mass is so small that it becomes 
irrelevant for {\it practical purposes} of the LHC physics.
It is also argued in \cite{nason} 
 that top quark mass measurements that provide 
the  most precise values of the top quark mass and that are based 
on the reconstruction of  kinematics of  top quark decay products 
should be thought of as 
measurements of the pole mass. The issue to address 
is not which mass parameter is measured but whether or not 
all sources of systematic uncertainties, including 
proper treatment of non-perturbative effects,  are accounted for.

%%%%%%%%%%%%%%%%%%%%%%%%%%%%%%%%%%%%%%%%%%%%%%%%%%%%%%%%%%%%%%%%%%%%%%%%%
%%
%%   use this format to include an .eps figure into your paper
%%
\begin{figure}[t]
\centering
\includegraphics[trim = 12 300 15 250, clip,  width=.7\linewidth]{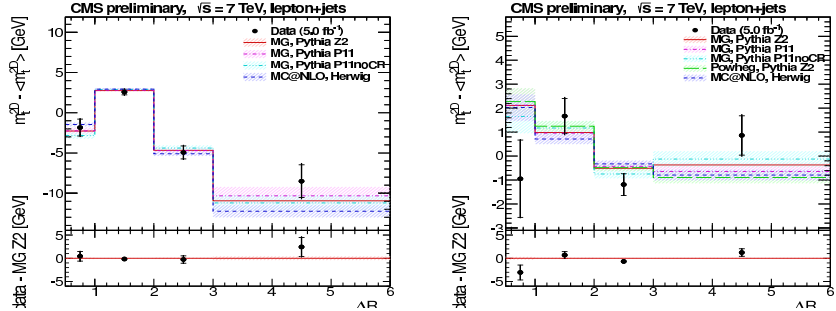}
\caption{Dependence of the extracted value of the top quark mass on applied kinematic 
cuts.}
\label{fig:fig1a}
\end{figure}
%%%%%%%%%%%%%%%%%%%%%%%%%%%%%%%%%%%%%%%%%%%%%%%%%%%%%%%%%%%%%%%%%%%%%%%%%%%

It is interesting to mention that there is a way to verify this  
point to some extent. Indeed, there are early 
studies by the CMS collaboration that explore  the dependence of the 
measured mass parameter  on the event  selection  criteria, c.f. Fig.~\ref{fig:fig1a}.
The idea is that by  doing so one can detect missed or poorly 
understood  non-perturbative 
effects since they should affect different events samples in different ways.  
It was found that the measured value of the top quark mass is stable against 
modifications of kinematic cuts.  However, 
the uncertainties in  the top quark mass reported in that study were   close to 
${\cal O}(2)$~GeV and similar studies with a much 
higher precision should be performed in the future.

\section{Simple processes} 

Theoretical description of processes where either a $t \bar t$ pair 
or a single top quark are produced are extremely advanced. Consider as an example 
the $t \bar t$ pair production at the LHC.  The landmark computations 
of NNLO QCD corrections to this process reviewed in Ref.~\cite{papanastasiou}
were recently combined with NLO electroweak corrections
\cite{pagani}.  There are numerous 
phenomenological applications  of these computations -- from constraints 
on  parton distributions functions, to the determination of top quark 
pole mass and the strong coupling constant, to a hunt for broad(ish) resonances 
that decay to top quarks  and interfere with continuum top pair production, 
to the exclusion of elusive stops in the so-called stealth mass region. 

One of the questions that seems to be heavily discussed in connection 
with these computations is the ``right'' choice of the renormalization 
and factorization scales ~\cite{papanastasiou} where what is right 
and what is wrong 
is decided by an accelerated  convergence of the perturbative expansion 
for a particular observable. It is found, for example, that 
$\mu = M_T/2$ is an appropriate scale for certain transverse momenta 
distributions whereas $H_T/4$ is a better choice for other 
observables~\cite{papanastasiou}.  While these discussions are interesting, to my taste they are 
missing an  underlying physics picture that should guide 
choices of scales in situations where NNLO 
calculations {\it are not} available. 
Clearly, a physical  way to define proper scales must depend on kinematics 
and parton composition of a  particular 
event.  Approaches based on these considerations 
are  well-known and are  used 
e.g. in parton showers where  strong coupling constants 
are evaluated at scales that correspond  to daughter's  transverse momenta 
in  sequences of 
$1 \to 2$ branchings. It is also employed in MLM, CKKW and MINLO approaches
that combine parton showers with matrix elements calculations. These 
connections between scales and kinematics 
are based on physics  principles  that theorists  were able to understand so far 
and that, as a matter of fact, 
should be applicable in a broader context. Hopefully, 
studies of scale setting prescriptions described in \cite{papanastasiou}
will either lead to a confirmation of this physical picture or to its 
extension and refinement.

\begin{figure}[t]
\centering
\includegraphics[trim = 12 300 15 250, clip,  width=.9\linewidth]{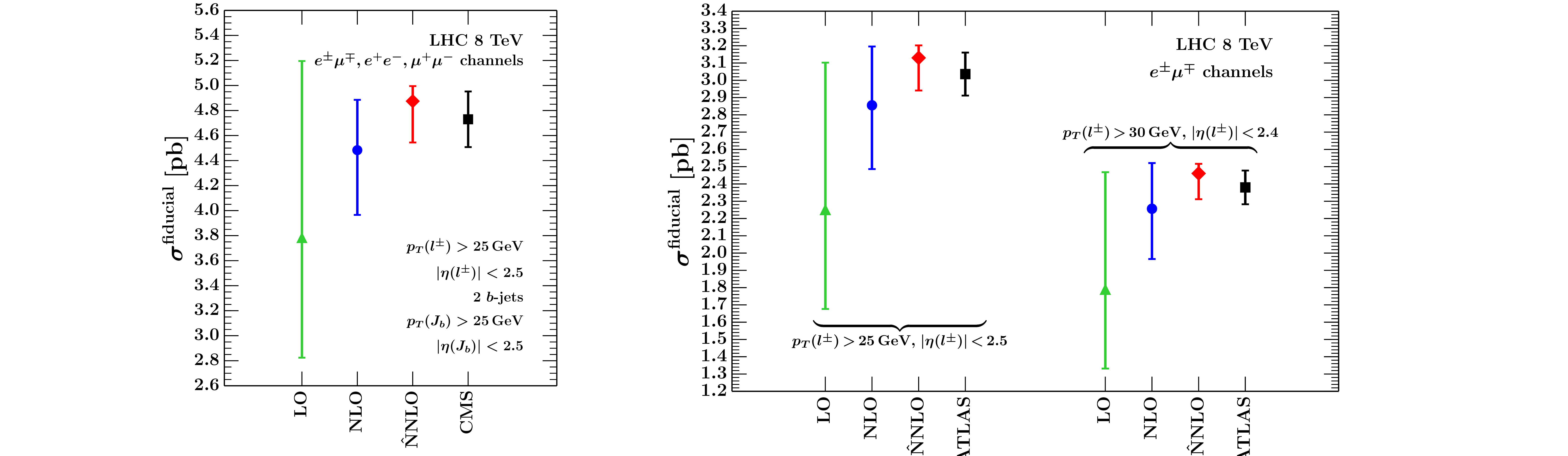}
\caption{Fiducial cross sections computed using approximate 
NNLO for production and exact NNLO for decay \cite{papanastasiou}. }
\label{fig:fig2}
\end{figure}

An important   next step in improving  description 
of top quark pair production 
is the inclusion of top quark decays; this will  enable computations 
of fiducial  cross sections defined at the level of physical particles 
and jets.   This can be done using  the narrow width 
approximation  that was well-tested in NLO QCD computations. 
The technology to combine top quark production and decay with higher-order 
QCD computations  is available  and 
it is only a matter of time before these results will be produced. Once 
this happens, there will be a tool to directly 
confront theoretical predictions 
with experimental  measurements and to study 
kinematic distributions of top quark decay products at the highest available 
level of precision.  Recall that similar computations at NLO QCD allowed 
for precision  studies of various  spin 
observables in top quark decays at higher orders 
of QCD, led to constraints on possible stop contributions to cross sections 
and provided important theory input for estimating how well the top quark 
mass can be inferred from kinematic distributions of top quark decay products. 
Once the NNLO production and NNLO decays are combined, 
it will be possible to perform similar studies at the next level 
of precision.   Although such results are not there yet, 
at TOP 2017  A.~Papanastasiou described  a computation 
\cite{papanastasiou} that combines approximate NNLO in the production 
with exact NNLO in the decay and it appears that even this approximate 
result  improves the description of fiducial 
cross sections dramatically, see   Fig.~\ref{fig:fig2}.

%%%%%%%%%%%%%%%%%%%%%%%%%%%%%%%%%%%%%%%%%%%%%%%%%%%%%%%%%%%%%%%%%%%%%%%%%
%%
%%   use this format to include an .eps figure into your paper
%%
%%%%%%%%%%%%%%%%%%%%%%%%%%%%%%%%%%%%%%%%%%%%%%%%%%%%%%%%%%%%%%%%%%%%%%%%%%%

Clearly, inclusion of QCD corrections to top decays enables direct 
computations of fiducial cross sections. This is good 
 but is it   worth the effort? Indeed, 
is it likely that the  QCD corrections to 
fiducial  cross sections differ significantly 
from corrections to total cross sections?  This question has 
the positive  answer   as the recent computation of NNLO QCD effects 
to single top quark production and decay shows \cite{gao}.
Both NLO and NNLO QCD corrections
to  inclusive single top production cross section are known 
to be small, close to one percent. 
However, if one computes corrections  to fiducial 
cross section that is defined 
by requiring  two jets, one of which should be 
a $b$-jet, with  $p_{\perp ,j} > 40~{\rm GeV}$ and $|\eta_j| < 5$, 
and imposes some constraints on leptonic transverse momenta and rapidities, 
QCD corrections to cross sections become significant and reach 
${\cal O}(-30\%)$ at NLO and ${\cal O}(-5\%)$ at NNLO. 
One of the interesting applications of the single 
top quark production is a 
constraint on the ratio of up and down PDFs that can 
be obtained from the ratio of single 
top and single anti-top cross sections or  kinematic distributions. 
As discussed in \cite{gao}
this ratio is stable  against radiative  corrections 
also for fiducial cuts,  
and seems to be predictable
with high,  ${\cal O}(1\%)$,  precision  at NNLO. 

Another interesting recent development is the appearance of 
NLO QCD computations that go {\it beyond} 
the narrow width approximation~\cite{papanastasiou}. Such 
computations  work with physical final states 
and include both resonance and non-resonance 
contributions and their interferences. The very appearance 
of these computations is a testimony to  a great  progress that theorists 
achieved in  computing  radiative corrections to hard scattering processes; none 
of these computations would have been 
 even remotely possible a decade ago, before 
the so-called NLO QCD revolution. 

What is the virtue of the calculations that go beyond the narrow 
width approximation?  I believe that quite often misleading 
answers to this question are given in that 
these calculations are attempted to be justified 
by arguing that there may be significant off-shell effects. 
As a rule, none are found, at least in kinematic  regions associated 
with  the $t \bar t$ production. 
Clearly, it   can not be otherwise  if 
 the top quark  samples are defined properly. Large off-shell effects 
are found in regions where production of two on-shell tops is kinematically 
unfavorable;  to understand what these regions are, no NLO QCD computations 
are needed since the analysis can be done at leading order. 
This is exactly  what experimentalists do when 
they design their selection cuts to  
obtain samples enriched with top quarks and this is why off-shell 
effects in  top quark production cross sections and distributions  are, typically,  small. 

Nevertheless, computations that  go beyond 
the narrow width approximation are very valuable 
because  they remove unphysical 
objects -- top quarks -- from the consideration 
 and allow us to define ``top  quarks'' operationally using kinematics of 
observable particles and jets  and selection cuts. It is this feature 
that changes the quality of  theoretical predictions, makes them infinitely 
closer to the real  world and  leads to theoretical  computations 
that are capable of reproducing quite closely 
what is being done in experiments. 
These developments are further supported by an understanding of 
how to combine these computations with parton showers through 
the resonance-aware matching algorithm~\cite{jezo} that will definitely contribute 
to making them  accessible to 
experimentalists. 

\section{Top quark couplings}

Couplings of top quarks to gauge bosons and the Higgs boson 
in the Standard Model are completely fixed (except for the CKM 
matrix elements that are largely arbitrary).
However, if we consider the 
Standard Model as an effective low-energy approximation 
to a more complete  theory, the Standard Model Lagrangian can be extended by 
a large number of higher-dimensional 
operators that may 
affect couplings of top quarks to other Standard Model particles.

%%%%%%%%%%%%%%%%%%%%%%%%%%%%%%%%%%%%%%%%%%%%%%%%%%%%%%%%%%%%%%%%%%%%%%%%%
%%
%%   use this format to include an .eps figure into your paper
%%
\begin{figure}[t]
\centering
\includegraphics[height=2in]{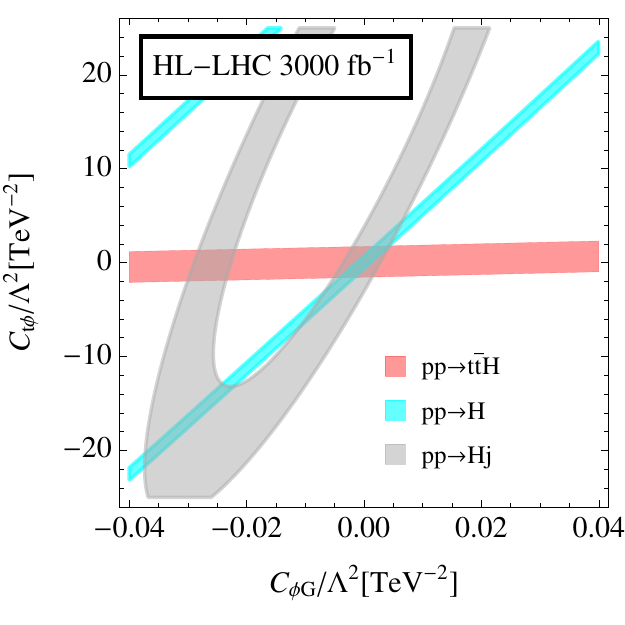}
\quad \quad 
\includegraphics[height=2in,width=2in]{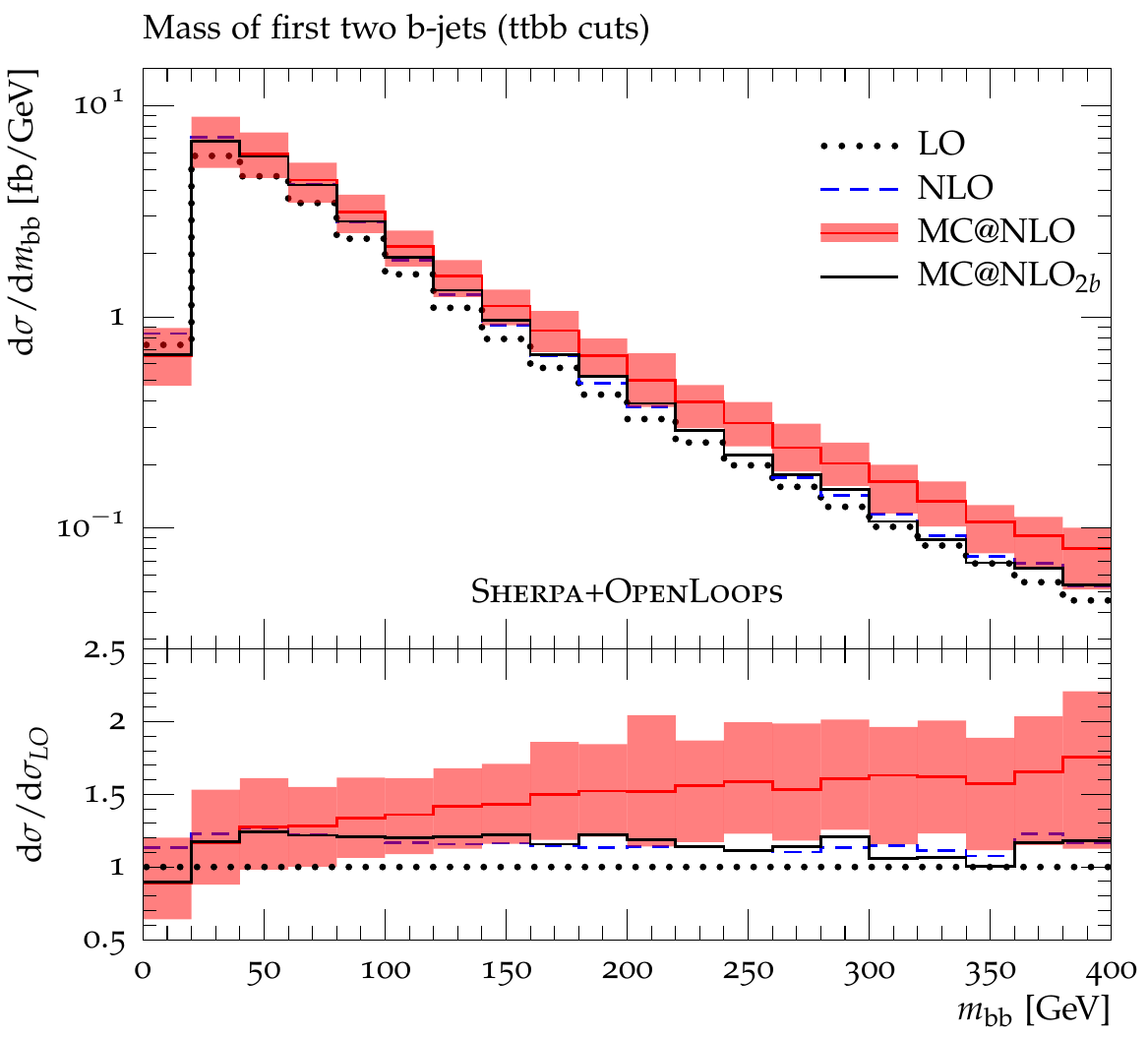}
\caption{
Left pane: expected constraints on the Wilson coefficients 
of the two operators in the extension of the Standard Model 
shown in Eq.(\ref{eq1}).  
See~\cite{vryonidou} for details. Right pane: 
Comparison of fixed order (NLO)  and matched predictions 
for the $b \bar b$ invariant mass spectrum in $pp \to t \bar t b \bar b$, 
see \cite{schumann}. }
\label{fig:fig5}
\end{figure}
%%%%%%%%%%%%%%%%%%%%%%%%%%%%%%%%%%%%%%%%%%%%%%%%%%%%%%%%%%%%%%%%%%%%%%%%%%%

Similar to precision electroweak or CKM fits, one can attempt 
to determine the Wilson coefficients of the many operators that 
appear in  extensions of the Standard Model  by fitting 
their  predictions to available data \cite{zhang, moore,vryonidou}.  
This is non-trivial 
since the parameter space is  large and the development of 
a careful strategy of how such a fit can be performed 
is required \cite{moore,schulze}.  While the discussion 
of the peculiarities of the full fit  is beyond the scope of 
this talk,  simple 
examples show that  
interesting things can be done. 
Indeed, imagine that  the Standard Model 
is extended by two operators that describe 
a local  $HGG$ interaction vertex and the 
modified top Yukawa  coupling
\begin{equation}
\label{eq1}
{\cal L}_{\rm SM} \to {\cal L}_{\rm SM} + \delta {\cal L},
\;\;\;\;
\delta {\cal L} = C_{\phi G} y_t^2 (\phi^+ \phi) G^{a}_{\mu \nu} G^{a,\mu \nu}
+ C_{t\phi} y_t^3 ( \phi^+ \phi) (\bar Q t) \tilde \phi.
\end{equation}  
The two  Wilson coefficients $C_{\phi G}, C_{t \phi}$ 
can be   over-constrained  \cite{vryonidou}
by studying contributions of effective operators 
in Eq.(\ref{eq1}) to three  
 production processes  that involve the Higgs boson and the 
top quarks, e.g.  
 $pp \to t \bar t H$,  
$pp \to H$ and $ pp \to Hj$. As follows from  the left pane in  
Fig.~\ref{fig:fig5}, 
this method allows a determination of the two Wilson 
coefficients with very high precision provided that 
$3000~{\rm fb}^{-1}$ of integrated 
luminosity are collected at the LHC.

Although  direct measurements of the top Yukawa coupling  
is of great interest,  it 
is a difficult measurement because of significant $t \bar t b \bar b$ 
background and a poor experimental resolution on the invariant mass of a $b \bar b$ 
pair.  It was expected that NLO QCD computations of $pp \to 
t \bar t b \bar b$  background would improve the quality of predictions 
for $b \bar b$ invariant mass distribution. Unfortunately, 
this did not happen since when  the NLO QCD computations 
were matched to parton showers, significant differences relative to fixed 
order results were  observed~\cite{schumann}. Indeed, gluon splittings
into $b \bar b$ pairs in parton showers
produce additional $b$-jets that increase  
the $m_{b \bar b}$ distribution 
in kinematic regions relevant for  top Yukawa  measurements by about 
thirty  percent. 
It it not clear at  this point  if 
this effect is real or if it is an artifact of the matching.
Further studies of this problem  are definitely warranted.

Another way to overcome the challenge of measuring the top quark 
Yukawa coupling is  to  better understand 
the $pp \to t \bar t H$ {\it signal}. 
 Since NNLO QCD computations for 
$2 \to 3$ processes are currently out of reach, one has to resort to 
approximate methods. In particular, as discussed by A.~Ferroglia, 
it is possible to improve existing 
NLO QCD predictions for $pp \to t \bar t H$ process by considering the resummation 
of soft gluons through next-to-next-to-leading logarithmic accuracy~\cite{ferroglia}.

\begin{figure}[t]
\centering
\includegraphics[height=2in]{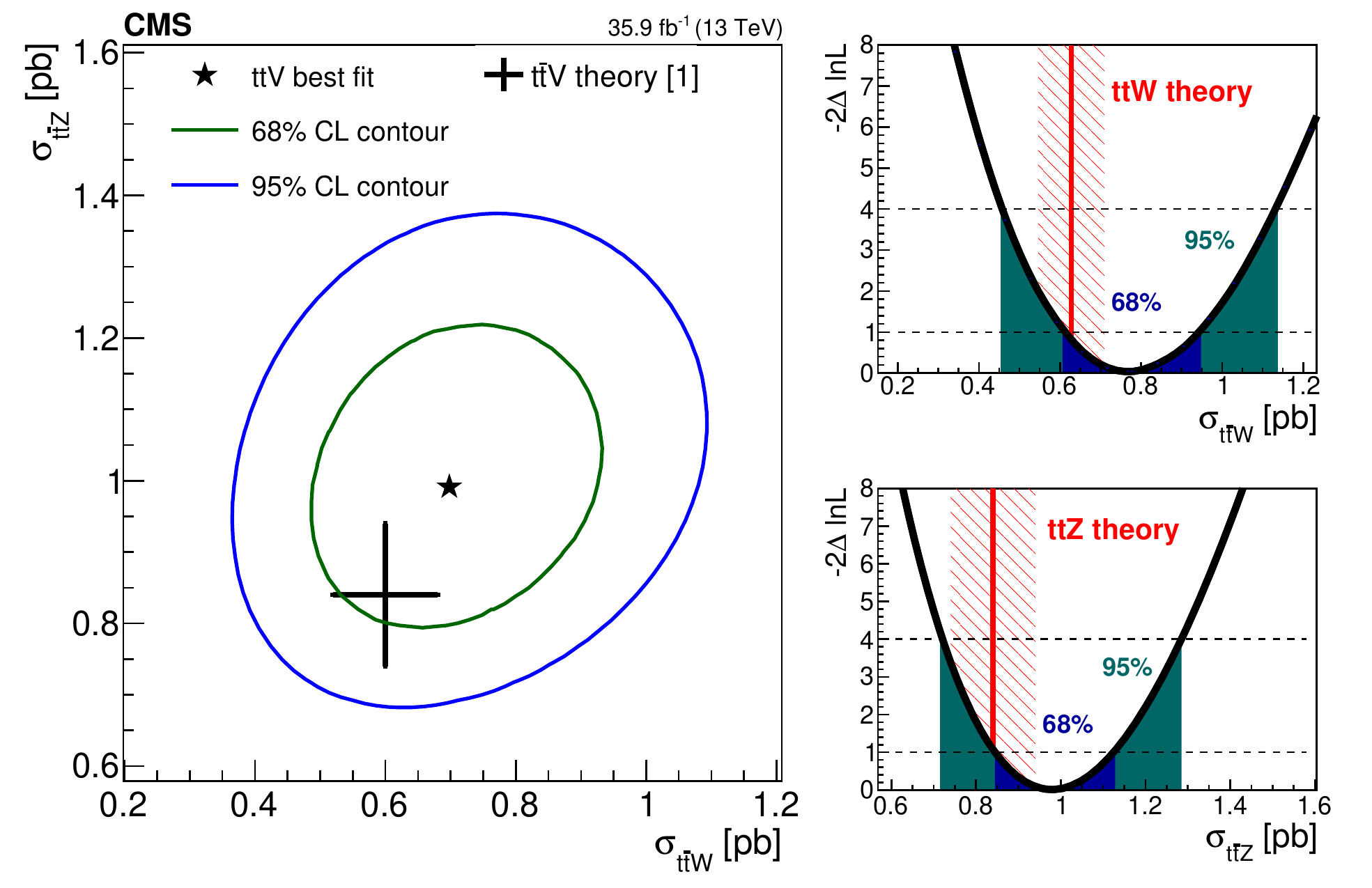}
\includegraphics[height=2in]{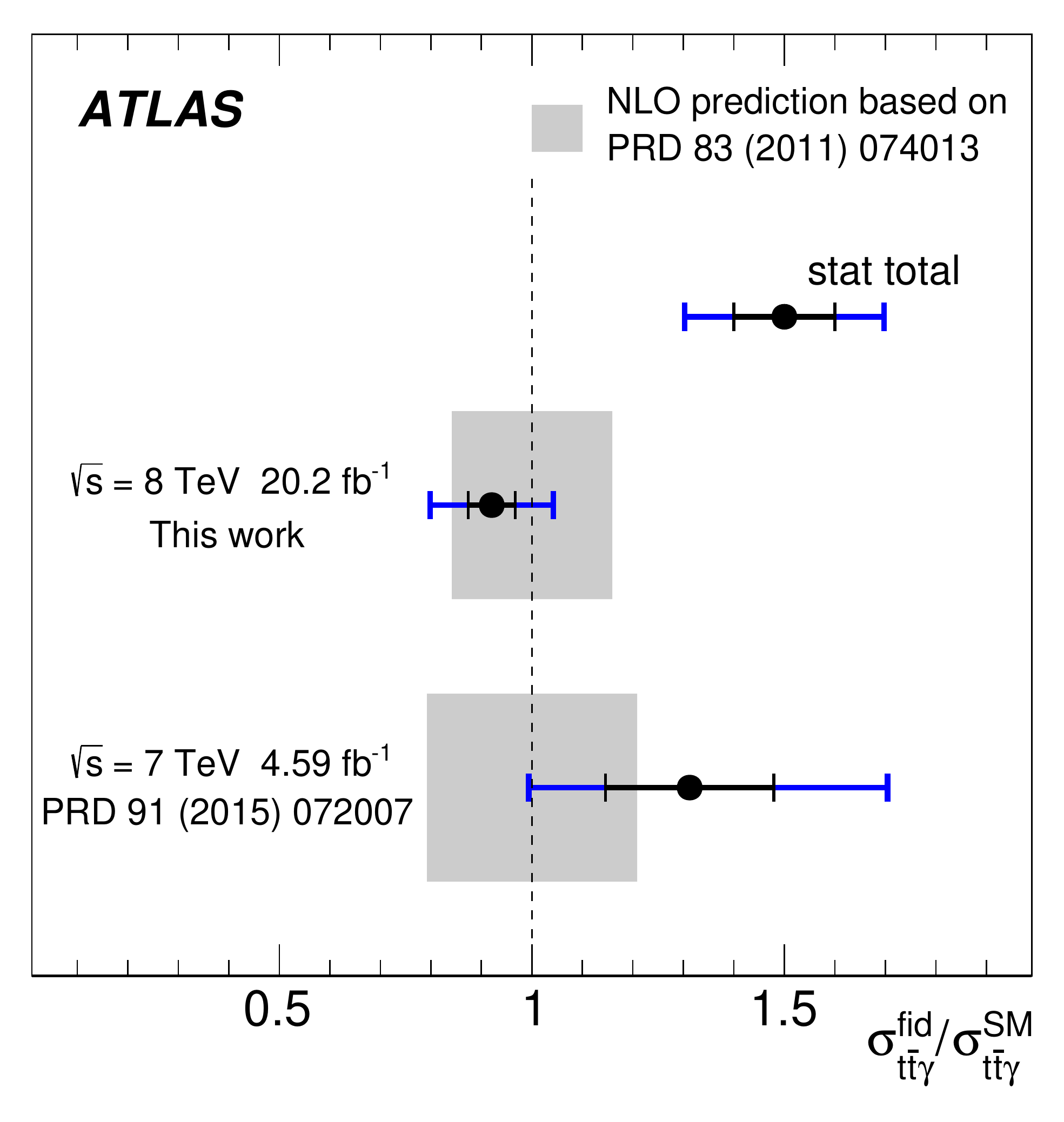}
\caption{
Comparison of measured $pp \to tt Z$, $pp \to t \bar t W$ 
and $pp \to t \bar t \gamma$ with theoretical predictions. }
\label{fig:fig4}
\end{figure}

Leaving the Yukawa coupling aside, 
it  is useful to keep in mind  that even  top quark couplings 
to electroweak vector bosons, including 
photons, are not well known  experimentally \cite{schulze}.
  For example,  there are still 
models consistent with experimental data 
that predict ${\cal O}(10\%)$ deviations  in $Zt \bar t$ couplings.
It is therefore exciting 
to see   that experimental measurements of $\sigma_{t \bar t Z}$
are starting to get 
close to this precision, see~Fig.~\ref{fig:fig4}.  
Precision, even marginal, requires  that NLO-corrected predictions 
for cross sections and kinematic distributions are used; such 
computations both in the Standard Model and allowing for 
more general structure of the $t \bar t Z$ couplings are discussed in  
\cite{schulze}.  Similar 
measurements of the $t \bar t \gamma$ couplings are also 
getting quite precise, see  Fig.~\ref{fig:fig4}. 
In this case it is 
essential to include proper theory predictions for photons emitted  both 
in top production and top decay processes  as they give similar  contributions 
to  fiducial cross section \cite{schulze}.

%%%%%%%%%%%%%%%%%%%%%%%%%%%%%%%%%%%%%%%%%%%%%%%%%%%%%%%%%%%%%%%%%%%%%%%%%
%%
%%   use this format to include an .eps figure into your paper
%%
%%%%%%%%%%%%%%%%%%%%%%%%%%%%%%%%%%%%%%%%%%%%%%%%%%%%%%%%%%%%%%%%%%%%%%%%%%%

\section{Conclusions}

If the future of hadron collider physics is in a stronger emphasis on precision 
-- as many null search results for physics beyond the Standard Model suggest -- 
the top  quark community  is posed to lead the  transition. Indeed,  
it is obvious that   already now top physics    
combines  precise measurements  with precise theory  and BSM insights in 
an impressive and  unique  way. 

Years of theoretical work  led to enormous   
progress that allows us to connect 
the top quark signals at the LHC 
with the Lagrangian  of the underlying quantum field theory, 
be it the Standard Model or one of its extensions. 
It is currently possible  to 
provide reliable  predictions for fiducial cross sections
that can be directly compared  with experimental measurements and 
use the results of these comparisons to  search   for physics 
beyond the Standard Model in a variety of clever  ways. Technology 
has been developed to study BSM physics in the top sector in an agnostic 
way that, hopefully, will guide us through the dark ages and 
enable New Physics discovery at the end.  The theory talks at the TOP~2017 
conference provided a powerful  testimony to just how far we got in our 
exploration of the top quark physics and charted a course for  the future. 
Without a doubt, there will 
a lot of interesting physics to discuss and summarize 
at the TOP~2018.  

\Acknowledgements
I would like to thank the organizers of the TOP 2017 for  enjoyable
and stimulating   conference.

\end{document}